\documentclass{article}
\usepackage{spconf,amsmath,graphicx,hyperref}
\usepackage{booktabs}
\usepackage{xcolor}
\usepackage{subcaption}

\title{VioPTT: Violin Technique-Aware Transcription from Synthetic Data Augmentation}

\name{Ting-Kang Wang$^{\star 2,3,1}$ \thanks{$^{\star}$ Equal contribution; Work done during internship at Sony CSL.} \qquad Yueh-Po Peng $^{\star 4,1}$ \qquad Li Su$^{2}$ \qquad Vincent K.M. Cheung$^{1}$ \qquad }

\address{
\normalsize $^{1}$ Sony Computer Science Laboratories, Inc., Tokyo, Japan \\
\normalsize $^{2}$ Institute of Information Science, Academia Sinica, Taipei, Taiwan \\
\normalsize $^{3}$ Graduate Institute of Communication Engineering, National Taiwan University, Taiwan \\
\normalsize $^{4}$ Original Content Center, Gamania Inc., Taipei, Taiwan \\
\small tim.tkwang@gmail.com, yabipeng@gamania.com, lisu@iis.sinica.edu.tw, cheung@csl.sony.co.jp\\
}

\begin{document}
\ninept
\maketitle
\begin{abstract}
% contributions
% - show that transfer learning is not necessary, that augmentation is helpful
% - end-to-end violin playing technique transcription
% - new augmentation pipeline for synthetic data creation
% - release synthetic playing technique dataset

While automatic music transcription is well-established in music information retrieval, most models are limited to transcribing pitch and timing information from audio, and thus omit crucial expressive and instrument-specific nuances. One example is playing technique on the violin, which affords its distinct palette of timbres for maximal emotional impact. Here, we propose \textbf{VioPTT} (Violin Playing Technique-aware Transcription), a lightweight cascade model that directly transcribes violin playing technique in addition to pitch onset and offset. Furthermore, we release \textbf{MOSA-VPT}, a novel, high-quality synthetic violin playing technique dataset to circumvent the need for manually labeled annotations. Leveraging this dataset, our model demonstrated strong generalization to real-world note-level violin technique recordings in addition to achieving state-of-the-art transcription performance. To our knowledge, VioPTT is the first to jointly combine violin transcription and playing technique prediction within a unified framework.
\end{abstract}

\begin{keywords}
articulation, music, playing technique, transcription, violin
\end{keywords}

\section{Introduction}
\label{sec:intro}
% The goal of automatic music transcription (AMT) is to convert an audio recording into a symbolic representation containing pitch and timing information. 
The goal of automatic music transcription (AMT) is to convert an audio
recording into a symbolic representation for music analysis and
performance understanding. As one of the most studied tasks in music information retrieval, AMT has achieved substantial progress with the advent of deep learning. In particular, instrument-specific models have been developed for piano~\cite{kong2021high, toyama2023automatic, hawthorne2017onsets, hawthorne2021sequence}, guitar~\cite{huang2023note, riley2024high}, and violin~\cite{tamer2023high}, while more recent approaches such as BasicPitch~\cite{2022_BittnerBRME_LightweightNoteTranscription_ICASSP} and MT3~\cite{gardner2021mt3} provide instrument-agnostic solutions. Despite these advances, most AMT systems are limited to note transcription and overlook expressive, instrument-specific performance nuances. 

For the violin, playing technique is integral to its rich timbral palette and emotional expressivity. Violin performance involves a wide range of fine-grained articulations delivered via different playing techniques (e.g., pizzicato, spiccato, flageolet) that are crucial to capturing its musical character. However, annotating these attributes is costly and requires expert knowledge, which has hindered the development of large-scale datasets beyond labeled single notes. As a result, existing violin transcription systems remain focused on pitch accuracy and timing, leaving articulation and playing technique largely unexplored. 

\begin{figure}
    \centering
    \includegraphics[width=.85\linewidth]{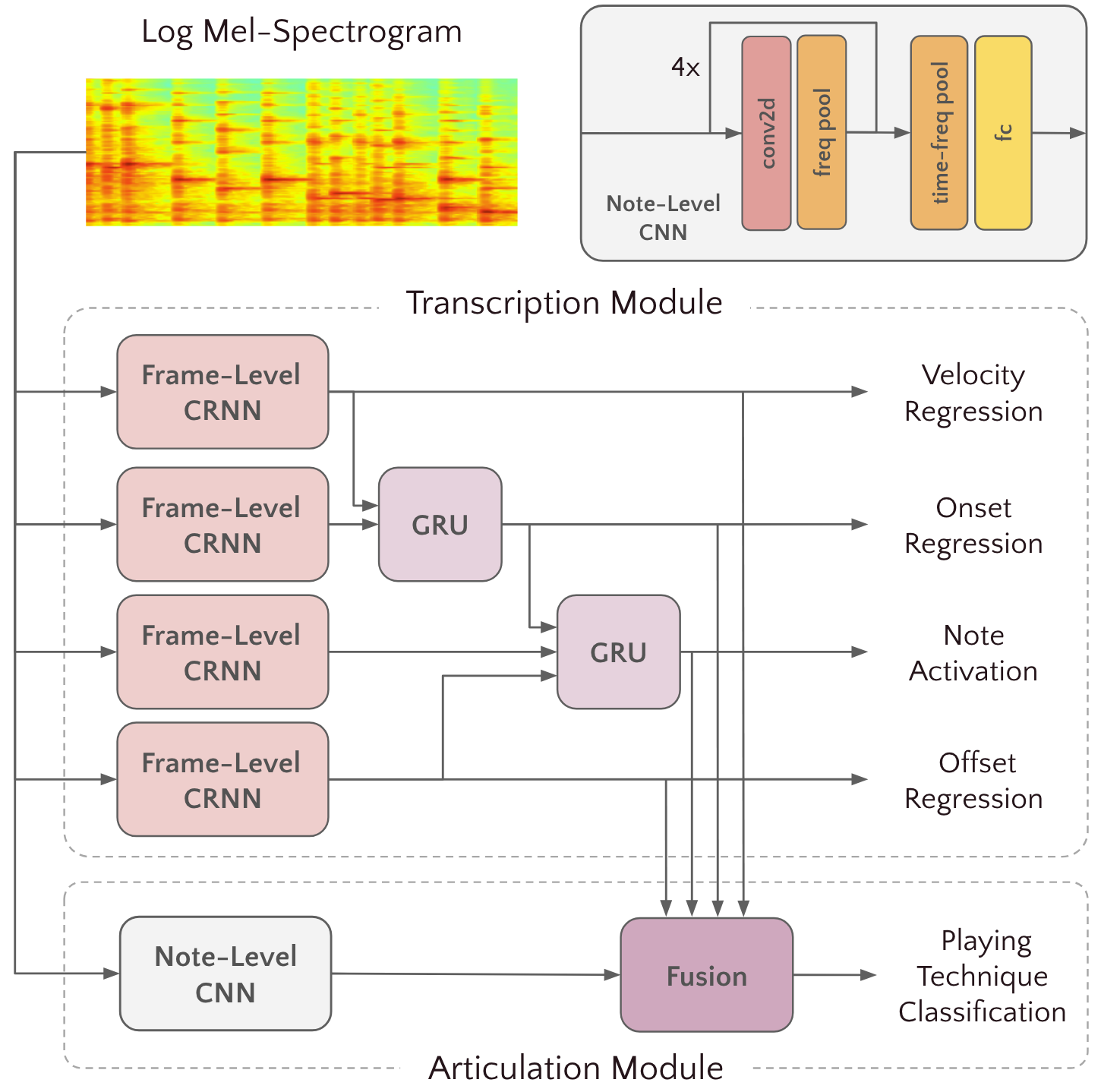}
    \caption{Overview of our technique-aware violin transcription model.}
    \label{fig:model-overview}
\end{figure}

To overcome these challenges, we propose a lightweight cascade violin transcription model that extends conventional AMT by jointly predicting pitch, onset, offset and playing technique. We show that state-of-the-art transcription performance is achievable by training from well-aligned solo-violin performances with augmentation and without relying on pretrained representations from other instruments. We further introduce and release a synthetic violin playing technique dataset to eliminate the need for expert technique annotations. Despite training on synthetic data, our model is able to successfully classify playing technique in real-world violin audio. Our approach thus provides a richer and more expressive representation of violin performance, and expand AMT towards capturing the full depth of musical expression. For reproducibility, model and code are available at \url{https://github.com/y10ab1/VioPTT}.

\section{Related Work}
Despite the recent emergence of instrument-agnostic transcription models (e.g., \cite{2022_BittnerBRME_LightweightNoteTranscription_ICASSP, gardner2021mt3}), piano transcription remains the most widely studied domain within AMT. The high-resolution model by Kong \textit{et al.}~\cite{kong2021high} achieved state-of-the-art performance on the MAESTRO dataset~\cite{hawthorne2018enabling} by modeling onset and offset times as regression targets rather than as discrete labels as in earlier work \cite{hawthorne2017onsets}. This enabled a precise localization of note boundaries beyond frame-level resolution using a light-weight recurrent neural network. By contrast, other recent models make use of the Transformer architecture to improve note timing and velocity prediction \cite{yan2021skipping, yan2024scoring, hawthorne2021sequence, toyama2023automatic}.

On the other hand, only one transcription model for violin currently exists. Tamer et al.~\cite{tamer2023high} introduced MUSC, a conformer-based system that transcribes fine-grained timing and pitch variations beyond the resolution of general-purpose AMT models. Their approach combined weak piece-level labels with iterative audio–score alignment to enable the modeling of pitch variations during \textit{glissando} and \textit{vibrato}. 

Similar to violin, guitar transcription is complicated by diverse playing techniques and timbral variations. Huang \textit{et al.}~\cite{huang2023note} proposed a guitar transcription model that combines a U-net encoder/decoder framework with multi-head attention to jointly predict note event and playing technique. By fine-tuning on Kong \textit{et al.}~\cite{kong2021high}'s transcription model originally trained on piano data, the guitar transcription model by Riley \textit{et al.}~\cite{riley2024high} achieved state-of-the-art performance and highlighted the value of domain adaption in AMT. Furthermore, instead of frame-level transcription as is typical in most AMT models, Kim \textit{et al.}~\cite{kim2022note} proposed a CNN-based note-level guitar transcription model via beat-informed quantization.

% Regarding violin technique classification, earlier work focused on modeling physical attributes of bow motion across different techniques \cite{maestre2010statistical}. Subsequent developments by Su \textit{et al.}~\cite{su2014sparse} demonstrated the feasibility of classifying playing technique in single notes using a support vector machine with sparse-coded magnitude and phase-derivative features from extracted from audio. Kruger \textit{et al.}~\cite{kruger2020playing} later expanded the model to allow simultaneous technique classification across violin, viola, cello, and double bass using augmented audio features extracted from a Hartley transform. A CNN-based violin technique classifier on single notes has also been recently proposed in \cite{alar2021audio}.

Recent work has also explored frame-level playing technique detection beyond note-based formulations. Li \textit{et al.}~\cite{li2023frame, li2022playing} proposed a frame-level multi-label approach with multi-scale convolution and self-attention to handle overlapping and mixed techniques in continuous performances. Building on this formulation, MERTech~\cite{li2024mertech} further incorporated self-supervised music representations and multi-task finetuning with auxiliary pitch and technique-onset prediction, achieving state-of-the-art results across multiple instrument technique datasets.

\section{Method}
\label{sec:method}

\subsection{Playing Technique-Aware Transcription Model}
\label{sec:model}
As illustrated in Fig.~\ref{fig:model-overview}, our model comprises two components: a \textit{transcription module} that predicts note information at the frame level, and an \textit{articulation module} that assigns a technique label to each transcribed note. The input is a log mel-spectrogram tensor with different STFT window lengths (length = $\{512,768,1024\}$, $f_s{=}16$\,kHz, hop size=$160$, mel-bins = $229$). 

We adopt Kong \textit{et al.}~\cite{kong2021high}'s high-resolution piano model for violin audio in the transcription module. One CRNN module is trained to separately predict at the frame-level onset, offset, velocity as regression targets, and note activation
 as a binary target. Each CRNN comprises four convolutional blocks followed by two bidirectional GRU layers. Features are then passed through a fully connected layer, two biGRUs, and a final 88-dimension sigmoid layer covering the violin pitch range.

The articulation module is designed to unify acoustic and transcription features for note-level technique classification. The input log mel-spectrogram tensor is encoded by four convolutional blocks (48, 64, 96, 128 channels) with pooling and dropout, followed by global average pooling and projection onto a 128-dimension acoustic embedding. In parallel, onset, offset, frame, and velocity features from the transciption module are projected to another 128-dimension embedding. The two embeddings are then concatenated and passed to a fully-connected layer in the fusion module, producing logits over 5 classes (4 techniques + no-technique).

% We design an articulation module that unifies acoustic and transcription features for note-level technique classification. The log mel-spectrogram is encoded by four convolutional blocks (48, 64, 96, 128 channels) with pooling and dropout, followed by global average pooling and projection to a 128-d acoustic embedding. In parallel, onset, offset, frame, and velocity features from a frozen transcriptor are projected to a 128 dim embedding. The two embeddings are concatenated and passed to a projection network, producing logits over 5 classes (4 techniques + non-technique), trained with cross-entropy loss.

The training objective is to minimize the sum of frame-wise onset, offset, velocity, and technique losses. In particular, binary cross-entropy is employed for frame, onset, and offset, mean squared error for velocity, and categorical cross-entropy for technique.

% \subsubsection{Joint Training Objective}
% \label{sec:objective}

% \begin{align}
% \mathcal{L} \;=\;
%    \ell_{\text{frame}}
%  + \ell_{\text{onset}}
%  + \ell_{\text{offset}}
%  + \ell_{\text{velocity}}
%  + \ell_{\text{tech}}. 
% \end{align}

% In particular, binary cross-entropy is employed for $\ell_{\text{frame}}$, $\ell_{\text{onset}}$, and $\ell_{\text{offset}}$, mean squared error for $\ell_{\text{velocity}}$, and categorical cross-entropy for $\ell_{\text{technique}}$.

% We minimize the sum of frame-wise, onset, offset, velocity, and technique losses:
% \begin{align}
% \mathcal{L} \;=\;
%    \ell_{\text{frame}}
%  + \ell_{\text{onset}}
%  + \ell_{\text{offset}}
%  + \ell_{\text{velocity}}
%  + \ell_{\text{tech}}, 
% \end{align}

% where Binary cross-entropy is employed for $\ell_{\text{frm}}$, $\ell_{\text{onset}}$, and $\ell_{\text{offset}}$, as these tasks are naturally formulated as probabilistic classification of note activity and temporal boundaries. Mean squared error is used for $\ell_{\text{velocity}}$ to regress continuous dynamics, while categorical cross-entropy is applied to $\ell_{\text{technique}}$.

\subsection{Data augmentation}

We considered two data augmentation strategies on (i) pitch and timing information, as well as (ii) playing technique synthesis.

\subsubsection{Semitone-Level Pitch and Timing Transcription}
\label{sec:augmentation strategies}
Using a customized effects chain inspired by guitar and synthesizer processing \cite{riley2024high}, we considered applying pitch and timing augmentation to training audio to enhance model robustness and mitigate overfitting. This consisted of pitch shifting ($\pm$0.1 semitones), a gain boost of +5~dB, two randomized band-pass filters with cutoff frequencies uniformly sampled between 32~Hz and 4096~Hz with variable resonance, and reverberation with a moderate room size parameter (0.35).

\subsubsection{Playing Technique Synthesis}
\label{sec:synthized dataset pipeline}
To obtaining note- and technique-aligned data without costly expert technique annotation, we synthesize a large-scale, annotation-free violin technique corpus directly from MIDI scores. Audio is rendered using \texttt{DAWDreamer}~\cite{Braun_DawDreamer_Bridging_the}, a Python audio processing framework that hosts VST instruments and a professional-grade virtual instrument, \emph{Synchron Solo Violin I}. We automatically control its key switches and continuous controllers (CCs) to render violin performance with different playing techniques, including \textit{détaché}, \textit{flageolet}, \textit{spiccato}, and \textit{pizzicato}. To minimize domain bias, we disable all room and spatial processing in the plugin and render mono stems at 16~kHz. As this pipeline is entirely annotation-free, it can be generalized beyond violin to any instrument with a VST to provide articulation control.

\section{Experiments}

We conducted two experiments to evaluate transcription performance of our model. First, we examined the impact of transfer learning and data augmentation (see Section~\ref{sec:augmentation strategies})
on transcribing pitch and timing information. To test whether knowledge from a related domain could benefit violin transcription, we initialized the transcription module with weights released by Kong et al.~\cite{kong2021high} pretrained on piano audio in the MAESTRO~\cite{hawthorne2018enabling} dataset. This yielded four conditions: with or without piano pretraining, and with or without pitch and timing augmentation. Second, we assessed our model's ability to generalize from synthetic data and classify violin playing technique in real-world data. We further performed an ablation study to examine the impact of each transcription feature on classification performance.

\subsection{Datasets}
We used the following datasets for model training:

\vspace{0.5em}

\noindent\textbf{MOSA \cite{huang2024mosa}.} MOSA(Music mOtion with Semantic Annotation) comprises 19 hours of professionally recorded solo-violin performances by 15 expert players. Each recording provides high-quality stereo audio aligned to note-level annotations, including pitch, rhythmic structure, dynamics, articulation, and harmonic analysis. This dataset serves as the core of our training data for which subsequent augmentations are built upon.
\vspace{0.5em}

\noindent\textbf{MOSA-VPT.} Using the synthesis pipeline described in Section~\ref{sec:synthized dataset pipeline}, we construct MOSA-VPT, a large-scale synthetic dataset of violin playing techniques\footnote{Available at \url{https://doi.org/10.5281/zenodo.18295471}}. This dataset comprises 76 hours of audio-MIDI pairs balanced across four techniques: \textit{détaché}, \textit{flageolet}, \textit{spiccato}, and \textit{pizzicato}. We target these four techniques because they exhibit distinct acoustic signatures and cannot be trivially modeled by existing transcription approaches (e.g., tremolo may be approximated as rapid détaché). 
% This selection provides a focused evaluation set for studying expressive violin performance.
This corpus provides note- and technique-aligned supervision at scale and is used to train our technique-aware transcription model.

\vspace{1em}

\noindent We used the following external datasets for model evaluation:

\vspace{0.5em}

\noindent\textbf{URMP~\cite{li2018creating}.}
URMP (University of Rochester Multi-Modal Music Performance) dataset consists of 44 chamber music recordings covering a wide range of instruments. Pitch annotations were first generated using pYIN~\cite{mauch2014pyin} through Tony~\cite{mauch2015computer}, and subsequently refined by manual correction of note onsets, offsets, and pitches. All violin tracks are used during test.

\vspace{0.5em}

\noindent\textbf{Bach10~\cite{datasetbach10}.}
Bach10 dataset comprises 10 four-part chorales performed by violin, clarinet, tenor saxophone, and bassoon. Ground-truth $f_0$ annotations were estimated using YIN~\cite{mauch2014pyin} with subsequent manual refinement. Note onsets are precisely annotated, while offsets remain uncorrected for comparison with previous work (\cite{tamer2023high}).

\vspace{0.5em}

\noindent\textbf{RWC~\cite{goto2003rwc}.} 
We use the Musical Instrument Sound subset of the RWC (Real World Computing) Music Database \cite{goto2003rwc} to evaluate real-world violin technique classification performance. This subset includes 50 instruments performing chromatic scales, each with three variations (such as different performers or manufacturers), under multiple dynamics and playing styles. We only use violin audio for testing, and separated the scale into single notes as in previous work.

% \vspace{0.5em}

% We use MOSA and MOSA-VPT as training corpora, 
% while the other datasets serve for external evaluation: 
% URMP and Bach10 for transcription, and RWC for playing technique assessment.
%% Explain why we choose these 4 playing techniques for our experiments.
% These four techniques are targeted because they exhibit distinct acoustic signatures 
% and cannot be trivially modeled by existing transcription approaches (e.g., tremolo often being approximated as rapid detaché). 
% This selection provides a focused evaluation set for studying expressive violin performance.

\subsection{Implementation}
The transcription module was trained for 10,000 steps with a batch size of 5, using 10-second violin recordings as input. The articulation module was trained separately for 1,000 steps with a batch size of 128, using 2-second single-note recordings. In both modules, samples were clipped or padded to match the target durations. All experiments were conducted on a single NVIDIA RTX~4090 GPU and we used a cosine-annealing learning rate scheduler with an initial learning rate of $5 \times 10^{-4}$.

For the transcription module, MOSA was used for training and validation. All performances of the first movement \textit{Preludio} from \textit{Bach’s Partita No.~3 in E major, BWV 1006} (about 10\% of MOSA) were held out for validation, and the remaining $\sim$90\% were used for training. The same piece with different performers was withheld to prevent data leakage. As with most transcription models, we omit velocity transcription and set it to a constant value.  For the articulation module, we used MOSA-VPT in its entirety as training data. The RWC dataset was divided into three folds for evaluation. In each run, two folds were used for validation and the remaining fold for test, and this procedure was repeated three times so that every fold served as the test set once. 

For pitch and timing transcription, performance was assessed using Precision (P), Recall (R), F1-score (F1), and onset-only F1-score (F1$_{\text{no}}$) in the \texttt{mir\_eval} library \cite{raffel2014mir_eval}. For P, R, and F1, a note is regarded as correct if its pitch deviation is within 50 cents, onset deviation is within 50 ms, and offset deviation is within 20\% of the note duration. For F1$_{\text{no}}$, the same criteria for F1 is applied without considering offsets. For technique transcription, performance was evaluated in two ways: First, using \textit{macro accuracy}, defined as the mean of accuracy values computed independently for each class to account for label imbalance; Second, using per-class accuracy. 

As baselines, we compared our work against the existing state-of-the-art violin transcription model MUSC~\cite{tamer2023high} and MERTech~\cite{li2024mertech}, a state-of-the-art multi-label technique classification model. As MERTech relies on technique-annotated data,  it was trained exclusively on the MOSA-VPT dataset following the original experimental setting for 200 epochs. Although MERTech is not explicitly designed for note transcription, pitch and onset information are incorporated as auxiliary cues and so we evaluated MERTech on both note and technique transcription.

\section{RESULTS AND DISCUSSION}
\label{sec:results}

\subsection{Pitch and Timing Transcription}
\label{sec:transcription result}

Table~\ref{tab:transcription_results} reports pitch and timing transcription performance with and without piano pretraining, as well as with and without data augmentation. On URMP, our model achieved the highest Recall (83.6) and F1\textsubscript{no} (93.1), while also matching the state-of-the-art MUSC~\cite{tamer2023high} in Precision and note-level F1 when trained from scratch and with data augmentation. On Bach10, the best performance across all metrics was achieved by our model trained with piano pretraining plus violin fine-tuning and without augmentation, closely followed by our model trained from scratch and with augmentation. These show that performance gains from transfer learning are limited when sufficient domain-specific data are available. We attribute this to differences in timbre and temporal characteristics between piano and violin, which may limit the transferability of pretrained acoustic representations. Furthermore, that our model exceeded MUSC performance despite being trained on $\sim$30\% less data also highlight the benefit of data augmentation and importance of well-aligned data.

Given these results, we use our model trained from scratch with pitch and timing augmentations as the transcription module in the next section.

\begin{table}[h!]
\centering
\resizebox{\columnwidth}{!}{
\renewcommand{\arraystretch}{1.1}
\setlength{\tabcolsep}{3.5pt}
\begin{tabular}{l|cccc|cccc}
\hline
\multicolumn{1}{c|}{} & \multicolumn{4}{c|}{\textbf{URMP}} & \multicolumn{4}{c}{\textbf{Bach10}} \\
\textbf{Model} & P & R & F1 & F1\textsubscript{no} & P & R & F1 & F1\textsubscript{no} \\
\hline

Ours w/o aug & 83.4 & 81.2 & 82.2 & 92.8 & 66.7 & 71.3 & 68.9 & 79.0 \\
Ours w/ aug & \underline{86.1} & \textbf{83.6} & \underline{84.5} & \textbf{93.1} & \underline{68.1} & \underline{71.8} & \underline{69.9} & \underline{79.5} \\
Ours + FT w/o aug & 84.4 & 79.0 & 81.3 & 91.3 & \textbf{69.5} & \textbf{73.7} & \textbf{71.5} & \textbf{80.2} \\
Ours + FT w/ aug & 85.0 & 82.1 & 83.3 & 92.9 & 63.3 & 68.4 & 65.7 & 77.8 \\
\hline
MUSC \cite{tamer2023high} & \textbf{86.5} & \underline{83.1} & \textbf{84.6} & \underline{93.0} & 65.0 & 64.8 & 64.8 & 77.0 \\
\textit{MERTech} \cite{li2024mertech} & 
26.6 & 33.7 & 29.8 & 30.3 &
27.6 & 53.4 & 36.4 & 36.9 \\
\hline
\end{tabular}
}
\caption{Transcription performance on the URMP and Bach10 violin tracks. ``FT" denotes models fine-tuned from a piano-pretrained checkpoint; ``aug" denotes augmentation. Bold and underlined values denote the best and second-best results per column, respectively.}
\label{tab:transcription_results}
\end{table}

% \begin{table*}[htbp]
% \centering
% \resizebox{.9\textwidth}{!}{
% \begin{tabular}{llllll}
% \toprule
%  & Macro Acc (\%) & Flageolet Acc (\%) & Détaché Acc (\%) & Pizzicato Acc (\%) & Spiccato Acc (\%) \\
% \midrule
% Full ablation & \underline{70.46} ($\pm$ 2.57) & \underline{86.44} ($\pm$ 4.19) & 51.75 ($\pm$ 9.97) & 57.06 ($\pm$ 15.33) & \textbf{86.56} ($\pm$ 2.55) \\
% Frame excluded     & 66.21 ($\pm$ 13.24) & 71.79 ($\pm$ 16.53) & \textbf{70.16} ($\pm$ 32.58) & 63.80 ($\pm$ 38.66) & 59.10 ($\pm$ 19.71) \\
% Offset excluded & 59.71 ($\pm$ 10.19) & 72.80 ($\pm$ 27.65) & 55.41 ($\pm$ 24.71) & 52.75 ($\pm$ 45.82) & 57.85 ($\pm$ 24.79) \\
% Onset excluded    & 65.82 ($\pm$ 8.63) & \textbf{91.55} ($\pm$ 1.96)  & 51.94 ($\pm$ 19.77) & \underline{65.47} ($\pm$ 20.35) & 54.34 ($\pm$ 11.90) \\
% Velocity excluded & 55.59 ($\pm$ 3.55) & 77.07 ($\pm$ 22.73) & \underline{65.67} ($\pm$ 26.70) & 0.16 ($\pm$ 0.28)   & 79.45 ($\pm$ 2.81) \\
% No ablation    & \textbf{77.22} ($\pm$ 6.35) & 71.89 ($\pm$ 14.12) & 63.12 ($\pm$ 12.59) & \textbf{88.80} ($\pm$ 3.11) & \underline{85.08} ($\pm$ 4.87) \\ 
% \bottomrule
% \end{tabular}
% } % end resizebox
% \caption{Comparing violin technique prediction performace on RWC \cite{goto2003rwc} with transcription features ablated. We report mean and standard deviation from three random data splits. Bold and underlined values respectively denote best and second-best results per column.}
% \label{tab:final_test_0915_3CV}
% \end{table*}

\begin{table*}[htbp]
\centering
\resizebox{.9\textwidth}{!}{
\begin{tabular}{llllll}
\toprule
 & Macro Acc (\%) & Flageolet Acc (\%) & D\'etach\'e Acc (\%) & Pizzicato Acc (\%) & Spiccato Acc (\%) \\
\midrule
Full ablation & \underline{70.46} ($\pm$ 2.57) & 86.44 ($\pm$ 4.19) & 51.75 ($\pm$ 9.97) & 57.06 ($\pm$ 15.33) & \textbf{86.56} ($\pm$ 2.55) \\
Frame excluded     & 66.21 ($\pm$ 13.24) & 71.79 ($\pm$ 16.53) & \textbf{70.16} ($\pm$ 32.58) & 63.80 ($\pm$ 38.66) & 59.10 ($\pm$ 19.71) \\
Offset excluded & 59.71 ($\pm$ 10.19) & 72.80 ($\pm$ 27.65) & 55.41 ($\pm$ 24.71) & 52.75 ($\pm$ 45.82) & 57.85 ($\pm$ 24.79) \\
Onset excluded    & 65.82 ($\pm$ 8.63) & \underline{91.55} ($\pm$ 1.96)  & 51.94 ($\pm$ 19.77) & \underline{65.47} ($\pm$ 20.35) & 54.34 ($\pm$ 11.90) \\
Velocity excluded & 55.59 ($\pm$ 3.55) & 77.07 ($\pm$ 22.73) & \underline{65.67} ($\pm$ 26.70) & 0.16 ($\pm$ 0.28)   & 79.45 ($\pm$ 2.81) \\
No ablation    & \textbf{77.22} ($\pm$ 6.35) & 71.89 ($\pm$ 14.12) & 63.12 ($\pm$ 12.59) & \textbf{88.80} ($\pm$ 3.11) & \underline{85.08} ($\pm$ 4.87) \\
% \midrule
% \textit{MERTech}$^{\dagger}$\cite{li2024mertech} & 5.43 & 10.55 & 9.91 & 0.87 & 0.39 \\
\midrule
\textit{MERTech} \cite{li2024mertech} & 53.36 ± (1.02) & \textbf{95.77} ± (2.23) & 58.80 ± (1.63) & 43.27 ± (1.19) & 15.61 ± (2.06) \\
\bottomrule
\end{tabular}
}
\caption{
Comparing violin technique prediction performace on RWC \cite{goto2003rwc} with transcription features ablated. We report mean and standard deviation from three random data splits. Bold and underlined values respectively denote best and second-best results per column.
% $^{\dagger}$MERTech results are reported under a technique-only prediction setting and are shown for reference.
}
\label{tab:final_test_0915_3CV}
\end{table*}

\subsection{Playing Technique Transcription}
\begin{figure}[htbp]
  \centering
  \includegraphics[width=0.8\columnwidth, trim=100 10 0 0, clip]{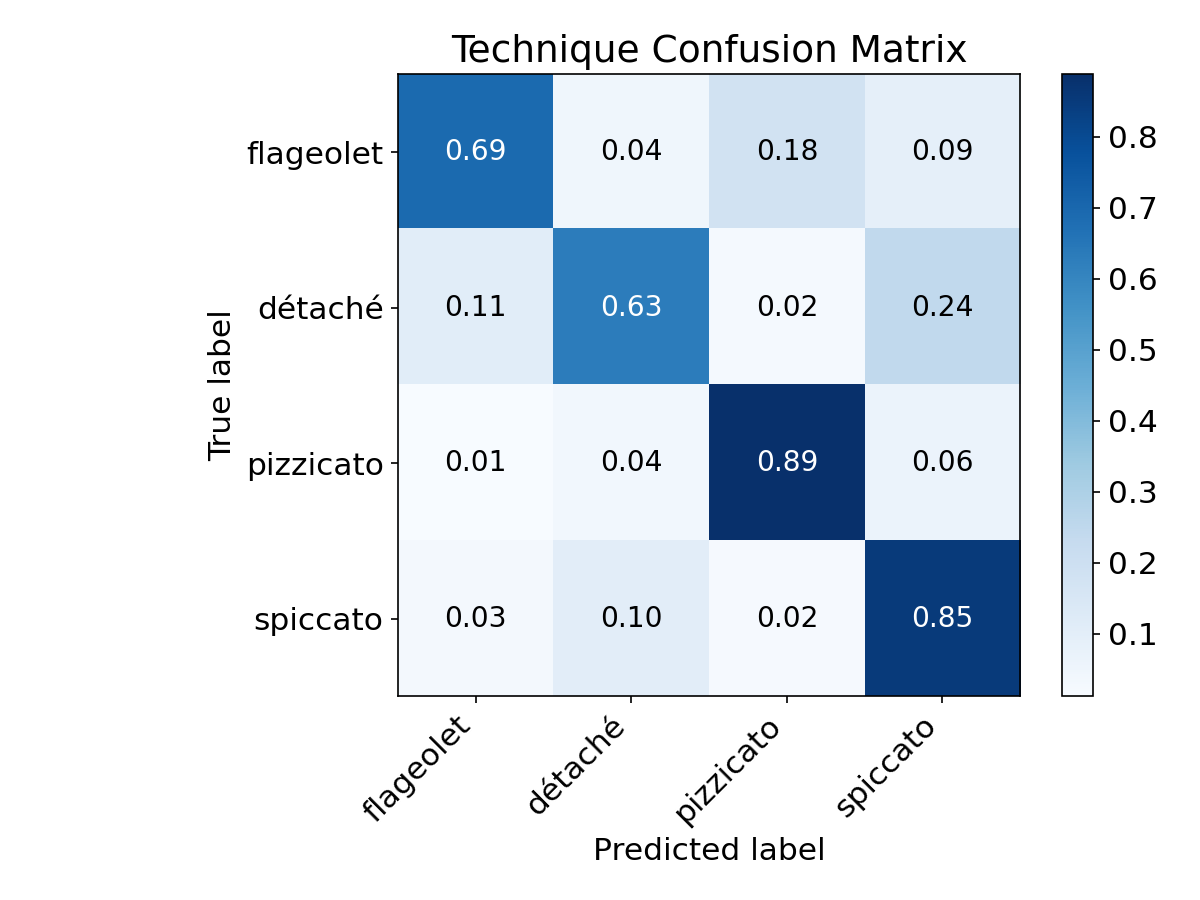}
  \caption{Confusion matrix for classification across four violin playing techniques using all transcribed features. Predictions were aggregated across all folds to highlight overall class-wise error patterns.}
  \label{fig:confusion_violin}
\end{figure}

Table~\ref{tab:final_test_0915_3CV} presents the violin technique transcription results on the RWC dataset under different transcription feature ablations in the articulation module. With the no ablation condition, our model achieved the highest macro accuracy of 77.22\%, confirming the benefit of having all transcription information. Removing individual features degraded overall performance: excluding offset led to a large drop to 59.71\%, while excluding frame or onset also reduced accuracy to a smaller extent. 

Examining each technique individually, MERTech~\cite{li2024mertech} achieved the highest accuracy for flageolet, with VioPTT obtaining the second-best performance when onset information was excluded. This suggests that our model primarily relied on harmonic features for flageolet classification. For détaché, accuracy improved when frame features were excluded. This indicates that binary frame information may have introduced noise, whereas offset and onset cues seemed to be more discriminative. Pizzicato showed strongest performance with no ablation, and excluding frame or onset only moderately reduced accuracy. However, excluding velocity proved to be detrimental and resulted in near-zero accuracy. This suggests that the model may have exploited velocity as a defining characteristic for this technique. Spiccato accuracy was similar without or with full ablation, with substantial performance drops when timing information was ablated. These indicate that classification relied on the combination of cues--especially timing information. Overall, the ablation study show that violin techniques exhibit distinct dependencies on transcription features: frame- and onset-based cues were particularly critical for techniques with sharp articulations such as détaché and spiccato, while harmonic cues dominated flageolet.

Similar conclusions can be reached when considering the confusion matrix of classification performance in the model without ablation in Fig~\ref{fig:confusion_violin}. While the model often misclassified détaché and spiccato, détaché was more often misclassified as spiccato than vice versa. This could be explained by spiccato classification relying on combination of features, although both techniques involve short bow strokes. On the other hand, low misclassification of pizzicato is consistent with its distinct acoustic and unbowed characteristic. 

\begin{figure}[htbp]
  \centering
  \includegraphics[width=0.70\columnwidth, trim=20 10 0 0, clip]{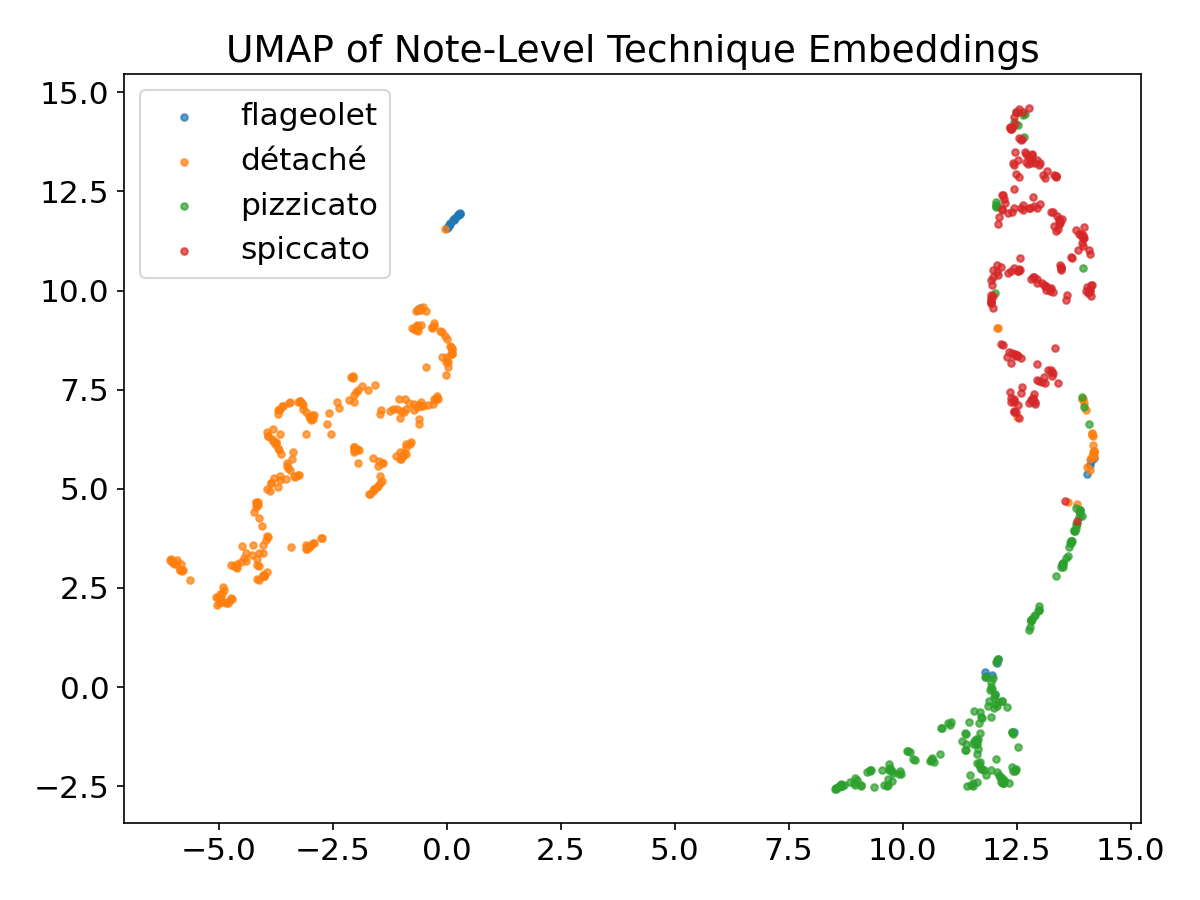}
  \caption{UMAP visualization of RWC data on learned note-level embeddings for four violin playing techniques.}
  \label{fig:umap_violin}
\end{figure}

Interestingly, a uniform manifold approximation and projection (UMAP) visualization of test data on the penultimate layer of the articulation module in Fig.~\ref{fig:umap_violin} suggests a divergence between synthetic and actual technique representations. The détaché cluster seemed to be closer to the flageolet cluster instead of spiccato, which seemed to be instead closer to the pizzicato cluster. Nevertheless, all four classes were largely disentangled, suggesting that the learned representations could be generalized from synthetic to unseen real data.

% These suggest that while synthetic data may be sufficient for technique-aware transcription, high-quality real-world data remain the gold standard.

% For the quantitative results reported in table \ref{tab:final_test_0915_3CV}, we computed performance metrics independently on each fold and report the mean and standard deviation across the three folds. This provides a fair comparison across folds and reflects the variability of the model under different data splits. For visualization, however, we constructed confusion matrices (see Fig~\ref{fig:confusion_violin}) by aggregating predictions from all folds and computing the statistics over the combined results. This yields a more comprehensive view of class-wise error patterns, while the reported mean~$\pm$~std values in the table remain the primary indicators of overall performance.

% \subsection{Latent Space Interpretation}
 % We present a uniform manifold approximation and projection (UMAP) visualization of the embeddings extracted from the penultimate layer of our articulation module. As the Fig.~\ref{fig:umap_violin} shows. The embeddings form well-separated clusters, indicating that the model captures discriminative technique-specific information. Flageolet shows acompact and isolated cluster, reflecting its distinctive harmonic timbre, whereas détaché is more dispersed, consistent with its continuous bowing variability. Pizzicato and spiccato clusters lie adjacent, aligning with their shared onset-driven and percussive acoustic characteristics. These separations demonstrate that the learned embeddings generalize to unseen real data.

\section{Conclusion and Future Work}
\label{sec:conclusion}
% While pitch and timing information remain fundamental in AMT, transcribing expressive nuances from different playing techniques is a crucial next step towards fine-grained music performance modeling. Here, we presented VioPTT, a light-weight cascade technique-aware transcription model for violin, and release MOSA-VPT, a high-quality synthetic violin playing technique dataset. In addition to exceeding the state-of-the-art in violin pitch and timing transcription, our model demonstrated high accuracy in transcribing playing technique in real-life performances despite training only from labeled synthetic data. This demonstrates the feasibility of using synthetic data as a proxy to technique annotations beyond single notes to obtain a richer and more nuanced transcription of expressive music performance. We envision our work to see applications in synthesis, performance analysis, and pedagogy. For future work, we aim to capitalize on our data synthesis approach to include more a wider variety techniques, as well as to extend the model to other bowed string instruments. 

While pitch and timing remain fundamental in AMT, modeling expressive nuances across playing techniques is essential for fine-grained performance transcription. We present VioPTT, a lightweight, technique-aware transcription model for violin, together with MOSA-VPT, a high-quality synthetic dataset for violin playing techniques. Our model achieves state-of-the-art performance in pitch and timing transcription and accurately recognizes playing techniques in real-world performances despite being trained only on synthetic data. This demonstrates the feasibility of using synthetic data as a proxy to technique annotations beyond single notes to obtain a richer and more nuanced transcription of expressive music performance. Future work will extend our synthesis framework to a broader range of techniques and other bowed string instruments.

\clearpage

\section{Acknowledgments}
This work was supported by funding from Sony Computer Science Laboratories, Inc. and the National Science and Technology Council, Taiwan under grant MOST 110-2221-E-001-010-MY3.

\bibliographystyle{IEEEbib}
\bibliography{strings,refs}

\end{document}